\documentclass[aps,prd,twocolumn,floatfix]{revtex4-1}
\usepackage{tabularx}
\usepackage{graphicx}
\usepackage{amsmath}
\usepackage{amssymb}
\usepackage{comment}
\usepackage{color}
\usepackage[svgnames]{xcolor}

\definecolor{mypink1}{rgb}{0.858, 0.188, 0.478}

\newcommand{\nruns}{N_{\rm runs}}
\newcommand{\niter}{N_{\rm iter}}
\newcommand{\pvalue}{$p$-value }

\newcommand{\reviseOne}[1]{#1}
\newcommand{\reviseTwo}[1]{#1}
\newcommand{\reviseThree}[1]{#1}
\newcommand{\reviseMinorTwo}[1]{#1}
\newcommand{\reviseMinorFive}[1]{#1}
\newcommand{\reviseMinorEight}[1]{#1}
\newcommand{\reviseMinorNine}[1]{#1}
\newcommand{\reviseMinorTen}[1]{#1}
\newcommand{\reviseNew}[1]{#1}

\begin{document}
\title{Towards a real-time fully-coherent all-sky search for gravitational waves from compact binary coalescences using particle swarm optimization}
\author{Marc E.~Normandin}
\affiliation{Dept. of Biology, University of Texas San Antonio, One UTSA Circle, San Antonio, TX 78249}
\author{Soumya D.~Mohanty}
\affiliation{Dept. of Physics and Astronomy, University of Texas Rio Grande Valley, One West University Blvd.,
Brownsville, Texas 78520}
\begin{abstract}

While  a fully-coherent all-sky search is known to be optimal for detecting gravitational wave  signals from compact binary coalescences, its high computational cost has limited current searches to less sensitive coincidence-based schemes. Following up on previous work that has demonstrated the effectiveness of Particle Swarm Optimization in reducing the computational cost of this search, we present an implementation that achieves near real-time computational speed.  This is achieved by combining the search efficiency of 
PSO with a \reviseOne{significantly revised and} optimized
numerical implementation of the underlying mathematical formalism
\reviseOne{along with additional multi-threaded} parallelization layers in
a distributed computing framework.  For
a network of four second-generation detectors with $60$~min data from each,
the runtime of the 
implementation presented here ranges between
$\approx 1.4$ to $\approx 0.5$ times the data duration for
 network signal-to-noise ratios (SNRs) of $\gtrsim 10$ and $\gtrsim 12$, respectively.
The reduced runtimes are obtained
with small to negligible losses in detection sensitivity: for  
a false alarm rate of $\simeq 1$~event per year in Gaussian stationary noise, the 
loss in detection probability is $\leq 5\%$ and $\leq 2\%$ for SNRs of 
$10$ and $12$, respectively. Using the fast implementation, we are able to quantify frequentist errors in parameter estimation for signals 
in the double neutron star mass range using a large
number of simulated data realizations. \reviseOne{A clear dependence of parameter estimation errors and detection sensitivity on the condition number of the network antenna pattern matrix is revealed. Combined with previous work, this paper securely establishes the effectiveness of PSO-based fully-coherent all-sky search across the entire binary inspiral mass range that is relevant to ground-based detectors.}
\end{abstract}
\maketitle

\section{Introduction} 
\label{sec:intro}
A network of spatially well-separated gravitational wave (GW) detectors is a critical 
requirement for GW astronomy. A detector network is essential for estimating the waveforms of the two polarizations of a GW signal and for localizing its source on the sky. 
Optimally combining the data from a detector network leads to improved GW search sensitivity. 

Starting with GW150914~\cite{PhysRevLett.116.061102}, a 
binary black hole inspiral and merger detected by the two-detector LIGO~\cite{2003SPIE.4856..282F} network, the LIGO-only and the 
LIGO-Virgo network collected $11$ confirmed compact binary coalescence (CBC) events
over  two observing runs~\cite{abbott2019gwtc}, O1 and O2.
The addition of Virgo~\cite{2013ASPC..467..151D} to the 
network played a
particularly important role in shrinking the localization error for GW170817~\cite{PhysRevLett.119.161101}, 
the coalescence of a double neutron star binary, leading to
the spectacular discovery of an electromagnetic counterpart~\cite{GW170817GWEM}. The LIGO-Virgo 
network is slated to be joined by the 
 Japanese KAGRA~\cite{2012CQGra..29l4007S} detector sometime during the ongoing (O3) observing run. In the near future, the planned worldwide network of five second-generation GW detectors will be completed with the construction and commissioning of LIGO-India~\cite{doi:10.1142/S0218271813410101}.

It is well-known that the optimal methods for  the
detection and estimation of CBC signals with network data~\cite{PhysRevD.64.042004} are the intimately related Maximum Likelihood Estimation (MLE)~\cite{kay_vol1} and Generalized Likelihood Ratio Test (GLRT)~\cite{1998.book.....KayII}, respectively. Both MLE and GLRT, conflated under the commonly used term fully-coherent all-sky search (FCAS)~\cite{Macleod_2016}, require the global
optimization of the joint likelihood function of data from a detector network over the full parameter space of CBC signals, which includes the
two sky angles, the masses of the binary components, and the components of their spins. 

While optimal, the computational cost of FCAS is daunting if the global optimization
is carried out over a regular grid in parameter space. The addition of a grid in the sky angles is estimated to increase the number of grid points by a factor of $O(10^3)$ 
over that for a single detector search~\cite{Macleod_2016}. This 
computational bottleneck has prevented an always-on FCAS search from being used on all of the 
data from a detector network. Instead, all search methods at present use
a semi-coherent scheme in which the data from each detector is first searched
separately and only those
 events 
 that pass a pair-wise coincidence test~\cite{PyCBC_GstLAL_2016PhRvD..93l2003A} are 
 followed up by FCAS search. The inability to
deploy FCAS search on all data has been estimated to result in a $25\%$
loss in the detection volume  for the first-generation LIGO-Virgo network~\cite{Macleod_2016}.

Even with the drastically reduced
live-time of the FCAS step in semi-coherent searches, grid-based optimization of the 
network likelihood for parameter estimation 
remains computationally infeasible. Instead, a  Markov Chain Monte Carlo (MCMC) based stochastic optimization approach~\cite{veitch2015parameter} is used to estimate
the parameters of candidate events. However, MCMC based methods are themselves  computationally expensive and slow, requiring another method called   \texttt{BAYESTAR}~\cite{singer2016rapid} that approximates the full MCMC to deliver rapid
sky localizations for electromagnetic follow ups. The speed of this method derives from using 
estimated values of parameters, other than the sky location, from the coincidence step. As such, it cannot serve as an FCAS search method.

Besides enhanced sensitivity, overcoming the computational barrier of an always-on FCAS search promises  other
potential advantages over semi-coherent searches. One is a simpler implementation that eliminates much of the empirical tuning based on ad hoc criteria that is involved in semi-coherent searches,  such as the tuning of per-detector 
detection thresholds and coincidence window size.
Another is that network analysis allows new kinds of vetoes~\cite{chatterji2006coherent} to be developed for non-astrophysical signals (``glitches"), further improving the sensitivity of an FCAS search.

It has been demonstrated in several studies by now that Particle Swarm 
Optimization (PSO)~\cite{PSO,bratton2007defining,engelbrecht2005fundamentals,mohanty2018swarm} offers a promising path forward in drastically reducing the computational cost of CBC searches.
 The first application of PSO to a GW data analysis problem in Ref.~\cite{Wang+Mohanty:2010} 
 demonstrated its effectiveness for a single-detector CBC search.
The application 
of PSO to FCAS search was proposed in Ref.~\cite{weerathunga2017performance} (WM) and,  for a network of first-generation detectors, showed a $10$-fold reduction in the number of likelihood evaluations compared to grid-based optimization. 
This prompted
further developments in Ref.~\cite{normandin2018particle} (NMW), where it was shown that 
the reduced computational burden of a PSO-based FCAS search also holds for data from
a second-generation detector network. 
In addition, a faster code was developed and improved convergence to the
global maximum was obtained by changing the variant of PSO used in the search.
An application of PSO to semi-coherent search itself~\cite{srivastava2018toward} has shown  
a large reduction in computational costs, further bolstering the
evidence for its effectiveness. 

In this paper, we present the next major step in the evolution of the PSO-based FCAS search: an optimized numerical implementation of the mathematical formalism 
combined with a multi-layered parallelized implementation that brings us to the doorstep of 
a real-time FCAS search. (By real-time, we mean a search that analyzes $T$~sec of data
in $T$~sec of wall-clock time.) The latest version of the code, called \texttt{BINARIES}  (Binary Inspiral Network Analysis Rapid Implementation Enabled by Swarm intelligence), can analyze $\approx 60$~min of data in $\approx 80$~min of wall-clock time for a target four-detector network signal-to-noise ratio (SNR) of $10.0$. 
The code becomes significantly
faster than real-time if the target is relaxed to ${\rm SNR}\gtrsim 12$ since
 the number of 
PSO iterations needed for the search are reduced considerably. 

\reviseOne{The efficient implementation of the mathematical formalism of FCAS search presented 
in this paper differs significantly from the one used in both WM and NMW. The parallelization layers are increased from two in NMW to three here through major changes, such as shifting to a multi-threaded numerical algorithms library. Together, these two developments make \texttt{BINARIES} $\approx 22$ times faster than the code used in NMW on the same computing hardware. }

Using \texttt{BINARIES}, we are able to obtain, for the first time,  
Frequentist error estimates for sky localization and 
chirp time parameters for the challenging case of a representative low mass 
($1.5 M_\odot$,$1.5 M_\odot$)  binary inspiral signal embedded in $60$~min of data. 
\reviseOne{This overcomes the limitations of WM and NMW to shorter signals 
and establishes the applicability of PSO-based FCAS search across the entire  mass range of binary inspirals relevant to ground-based detectors.}
\reviseOne{Using direct numerical estimation} allows a more realistic assessment of parameter estimation errors than
 analytic estimates  based on the Cramer-Rao
Lower Bound (CRLB)~\cite{kay_vol1} that is only attained asymptotically at high SNR. 
Further changes implemented in the present paper include the use of
detector-specific design sensitivity curves instead of the same, advanced LIGO, one for all.
The resulting error estimates are, therefore, relevant to 
the actual worldwide detector network.  

\reviseNew{The rest of the paper is organized as follows. Sec.~\ref{sec:fcas} provides a review of the FCAS search formalism, the noise and signal models used in this paper, and the changes made to the numerical implementation of the formalism. Sec.~\ref{sec:pso} discusses PSO and the tuning process used to optimize its performance. The runtime analysis of \texttt{BINARIES} is examined in Sec.~\ref{sec:comptime}. Results on the detection and estimation performance of \texttt{BINARIES} on simulated data are presented in Sec.~\ref{sec:performance}. 
We conclude with a discussion of the results and pointers to future investigations in Sec.~\ref{sec:discussion}.}
\section{Fully-coherent all-sky search}
\label{sec:fcas}
Much of the mathematical formalism for the fully-coherent 
all-sky search remains the same as in WM and NMW that, in turn, closely follow~\cite{PhysRevD.64.042004}. In this paper, we focus more on those aspects of the formalism
that were modified to improve the efficiency of its numerical implementation.

A $T$~sec long segment of data from
the $i$th detector in a network of $D$ detectors is denoted by
$x^i(t)$. Under the null ($H_0$) and alternative ($H_1$) hypotheses, 
\begin{eqnarray}
x^i(t) & = & n^i(t)\;,
\end{eqnarray}
and 
\begin{eqnarray}
x^i(t) & = & h^i(t) + n^i(t)\;,
\end{eqnarray}
respectively, where $n^i(t)$ is a noise realization and
$h^i(t)$ is the strain response of the detector to 
an incident GW signal. 

For a source located at azimuthal angle $\alpha$ and  polar angle $\delta$ in the Earth Centered Earth Fixed Frame (ECEF)~\cite{Leick_04}, the detector responses are given by,
\begin{eqnarray}
\left( \begin{array}{c}
h^1(t+\Delta^1(\alpha,\delta))\\
h^2(t+\Delta^2(\alpha,\delta))\\
\vdots\\
h^D(t+\Delta^D(\alpha,\delta))
\end{array}
\right) & = & {\bf F}(\alpha,\delta,\psi)\left(
\begin{array}{c}
h_+(t)\\
h_\times(t)
\end{array}
\right)\;,
\label{eq:detresponse}
\end{eqnarray}
where the ${i}^{\rm th}$ row of the antenna pattern matrix ${\bf F}(\alpha,\delta,\psi)$ contains the antenna pattern functions
$ (F_{+}^i(\alpha,\delta,\psi),\; F_{\times}^i (\alpha, \delta,\psi))$
of the ${i}^{\rm th}$ detector, $h_+(t)$ and $h_\times(t)$ are the TT gauge polarization components
of the GW plane wave 
incident on the origin of the ECEF, and $\Delta^i(\alpha,\delta)$ is the time delay between the plane wave hitting the ECEF origin and the ${i}^{\rm th}$ detector. The polarization angle $\psi$ gives the orientation of the wave frame axes with respect to 
the fiducial basis formed by  $-\widehat{\alpha}$ and $\widehat{\delta}$ in the plane orthogonal to the wave propagation direction. 

\subsection{Noise model}
\label{sec:noise_model}
\begin{figure}
\centering
\includegraphics[scale=0.21]{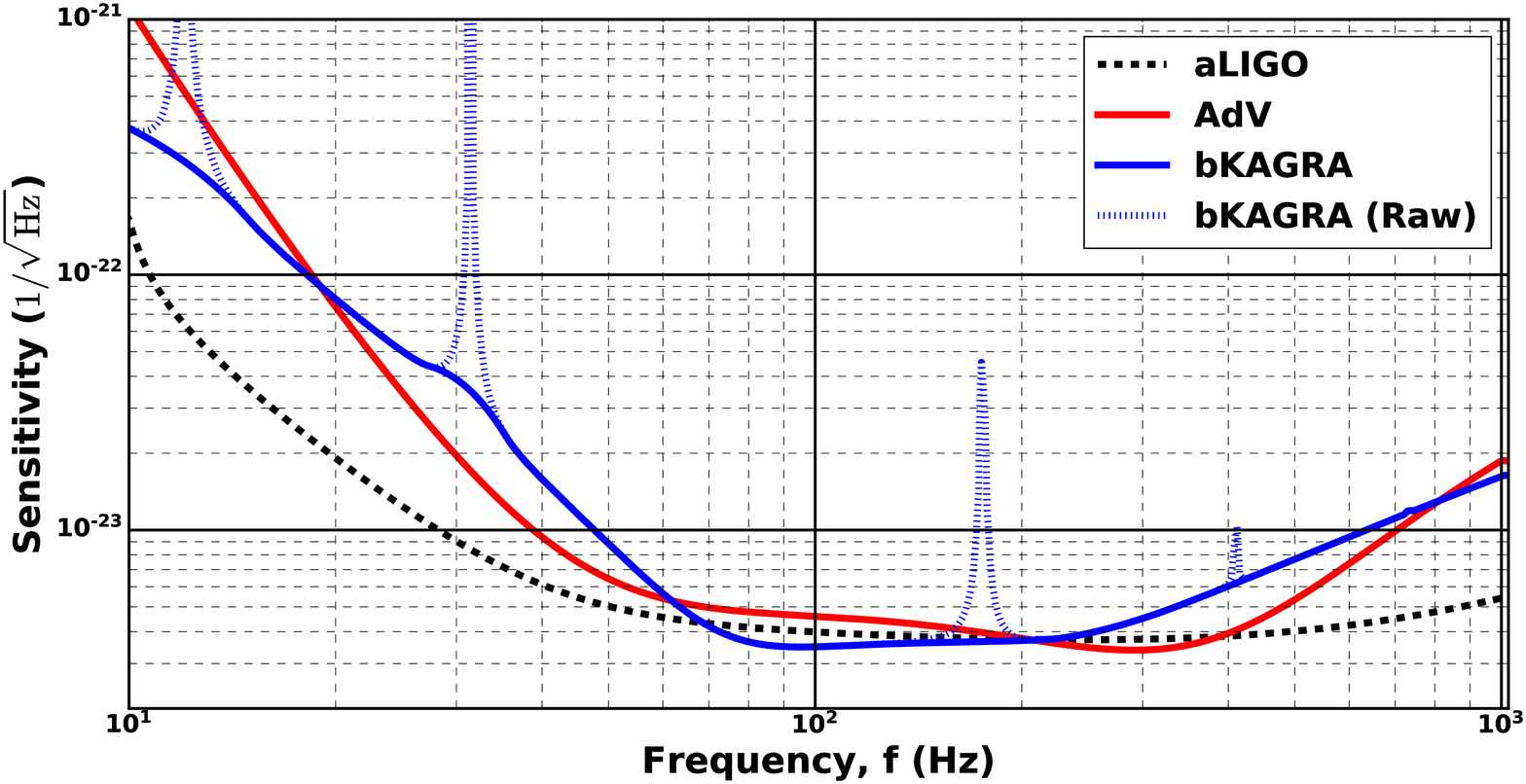}
\caption{\label{fig:detector_sensitivities}Strain sensitivity curves used in this paper
for aLIGO (Hanford and Livingston), advanced Virgo, and  
KAGRA.  The curves are labeled, respectively, as aLIGO, AdV, and bKAGRA.
The aLIGO curve is obtained from \cite{aligopsd}.
The bKAGRA interpolates across the lines in the actual
curve~\cite{lcgt_special_working_group_2009} labeled as bKAGRA (Raw) in the figure.
The AdV curve is obtained from \cite{avirgo_noise_curve}.}
\end{figure}

In common
with theoretical studies of detection and estimation performance of CBC search algorithms, we assume that  $n^i(t)$ is the 
realization of a stationary zero-mean Gaussian process with 
one-sided  
power spectral density (PSD) $S_n^i(f)$ at Fourier frequency $f$.   Further, $n^i(t)$
and $n^j(t)$, $i\neq j$, are assumed to be realizations of statistically independent stochastic processes. 
Figure~\ref{fig:detector_sensitivities} shows the PSDs used in this paper in the form of 
 strain sensitivity curves ($\sqrt{S_n^i(f)}$). These
 correspond to the design sensitivities of 
 the two aLIGO detectors at Hanford (H) and Livingston (L), advanced Virgo (V), and KAGRA (K).
 
Several high power narrowband noise features (``lines") are present in the \reviseMinorTwo{design sensitivity} of KAGRA.  \reviseMinorTwo{Due to
their adverse impact on the dynamic range of data and the associated numerical errors in its processing, the generation of simulated KAGRA noise must use a PSD that models the removal of these features.} However, 
without considering a specific line removal or whitening method, it is not possible to deduce
how much of the bandwidth associated with each line should be  notched or set to zero.
 In this paper, we follow the simple approach of interpolating the noise floor across each line (leaving behind a slight bump). Since the signal power in the corresponding bands is not suppressed to the same amount as the lines, we incur an overestimate of parameter estimation accuracy.  We leave it to future work to 
revisit this issue more carefully once
the characteristics of real KAGRA noise and specific line removal methods have been established. \reviseMinorTwo{It should be noted that lines are present in real data from all interferometic GW detectors and they are mitigated in the data conditioning step that precedes any analysis of real data. The line mitigation step in data conditioning for an FCAS search can be the same as the one used in a semi-coherent search.}

\subsection{Signal waveform}
The polarization waveforms used in this paper are obtained
from the restricted 2-PN formalism~\cite{Blanchet_95}
for a circularized binary with non-spinning components. In this paper, we only need to show the waveforms
schematically, with more detailed expressions available in, for example, WM.
Under the stationary phase approximation, the waveforms in the Fourier domain are
\begin{eqnarray}
    \widetilde{h}_{+}(f) &=& A_+f^{-7/6} \exp[{-i}\Psi(f)] \;,\\
    \widetilde{h}_{\times}(f) &=& A_\times   f^{-7/6} \exp[{-i}(\Psi(f)+\pi/2)]\;,\\
    \Psi(f) &=&  2\pi f t_c - {\phi_c} - \pi/4 + \psi(f)
\end{eqnarray}
 where $\psi(f)$ belongs to a two parameter family of smooth functions. The 
 parameters depend on the masses, $m_1$ and $m_2$, of the binary components but instead of using them directly, it is more convenient to use the chirp time parameters $\tau_0$ and $\tau_{1.5}$
 \begin{eqnarray}
\tau_{0} &=& \frac{5}{256\pi} f_{*}^{-1} \left(\frac{GM}{c^3}\pi  f_{*}\right)^{-5/3}\eta^{-1}\;,  \\
\tau_{1.5} &=& \frac{1}{8}f_{*}^{-1}\left(\frac{GM}{c^3}\pi  f_{*}\right)^{-2/3}\eta^{-1}\;, \\
M &=& m_1 + m_2 \;,\quad\mu = \frac{m_1m_2}{M}\;,\quad \eta =\frac{\mu}{M}\;,
\end{eqnarray}
 where 
 $f_\ast$ denotes the low-frequency cutoff of a high pass filter that must be applied to GW detector data before commencing any search in order to suppress the steep rise in $S_n^i(f)$ due to seismic noise. The effect of the high pass filter is taken into account by setting  
$\widetilde{h}_{+,\times}(f) = 0$ for $f\leq f_\ast$.  While 
$f_\ast$ is arguably detector-specific, we adopt a common value of $f_\ast = 10$~Hz in this 
paper for all second generation detectors. 

The other
parameters defining the waveforms are the overall amplitudes $A_{+,\times}$, which depend purely on the distance to the binary and its orientation relative to the line of sight;  the time, $t_c$, between the instantaneous frequency 
of the inspiral 
signal crossing $f_\ast$ and the plunge at the last
stable orbit that ends the inpiral;
the instantaneous 
phase, $\phi_c$, of the waveforms at $t_c$. 

Besides the 
low-frequency cutoff, $f_\ast$, all waveforms also have a high frequency 
cutoff caused by the plunge.
While this cutoff depends on the mass parameters of the system, the amount of time a low mass system spends
near plunge contributes very few cycles to the waveform relative to the inspiral phase,
allowing 
the waveform model to use a generic high frequency cutoff. This is set to 1000~Hz for the waveforms
considered in this paper.

\subsection{Detection and estimation}
For the noise model used in this paper, the log-likelihood ratio for a $D$ detector network is given by,
\begin{align} \lambda^{(D)} &= \sum_{i=1}^{D}\left[ {\langle}x^i|h^i{\rangle}^i - \frac{1}{2}{\langle}h^i|h^i{\rangle}^i\right]\;, \\
h^i & =  \sum_{k=1}^4 A_k h_k^i(t-\Delta^i)\;,
		\label{eq:loglikelihood}\\
{\langle}\;p\; |\; q\;{\rangle}^i  &= 4 \; {\rm Re} \int_{0}^{\infty} df\; \frac{\widetilde{p}(f)\widetilde{q}^\ast(f)} {{S_n^i}(f)}\;.\label{eq:innprod}
\end{align} 
Here, we have used the fact that $A_{+,\times}$, $\psi$,
and $\phi_c$ can be reparametrized as amplitudes, $A_k$, of the so-called template waveforms
\begin{eqnarray}
\begin{aligned}
{h_1^i}(t) &=& U_+^i h_c(t), \quad
{h_2^i}(t) &=& U_\times^i h_c(t), \\
{h_3^i}(t) &=& U_+^i h_s(t),\quad
{h_4^i}(t) &=& U_\times^i h_s(t),
\end{aligned}
\end{eqnarray}
where $U_a^i=F_a^i(\alpha,\delta,0)$, $a=+, \times$, and
$\widetilde{h}_c(f)=\widetilde{h}_+(f)$ ($\widetilde{h}_s(f)=\widetilde{h}_\times(f)$) for $\phi_c = 0$ and $A_{+,\times}=1$. 

Detection in FCAS is based on the GLRT  statistic, defined as
\begin{eqnarray}
   \rho_{\rm coh}^2 & = &\max_\Theta \Gamma^2(\Theta) \;,\label{eq:FCASdetStat}\\
  \Gamma^2(\Theta) & = &\max_{\Theta_{\rm ext}}
  \lambda^{(D)}\;,\label{eq:gammaSqd}
\end{eqnarray}
where $\Theta_{\rm ext}$ and $\Theta$ are the sets of so-called extrinsic and intrinsic parameters:
$\Theta_{\rm ext} = (t_c,\{A_k\})$, 
$k = 1,\ldots,4$, and 
$\Theta = (\alpha,\delta,\tau_0,\tau_{1.5})$. 
Maximization over $A_k$ can be carried out analytically
while $t_c$ can be efficiently maximized over using 
the Fast Fourier Transform (FFT). Maximization over
$\Theta$ must be done numerically and this dominates
the overall computational cost of the FCAS. We call $\Gamma^2(\Theta)$ the
{\em coherent fitness function} and $\rho_{\rm coh}$ the {\em coherent search statistic}.

The MLE estimates
of $\Theta$ and $\Theta_{\rm ext}$ are the global maximizers of the log-likelihood function. Since 
the log-likelihood differs from $\lambda^{(D)}$
only by a constant for given data,  the MLE estimates are obtained for 
free as part of the GLRT statistic calculation.

\subsection{Efficient evaluation of the log-likelihood ratio}
\label{sec:efficientNumerics}
After analytical maximization over $A_k$ in Eq.~(\ref{eq:gammaSqd}), one is left with the 
evaluation of 
 $\langle h^i_k(t-\Delta^i) | h^i_j(t-\Delta^i)\rangle$
and $\langle x^i | h^i_k(t-\Delta^i)\rangle$.
Each such inner product [cf. Eq.~(\ref{eq:innprod})] is implemented using the FFT
and involves the element-wise
product of two arrays followed by summation, leading 
to $O(N)$ floating point operations for data
segments containing $N$ samples. 
 Substantial savings in the
number of floating point operations can, therefore,
be obtained
by incorporating the following straightforward 
optimizations in the numerical evaluation of 
these inner products.

First, the inner product $\langle h^i_k(t-\Delta^i) | h^i_j(t-\Delta^i)\rangle$ is  
independent of $\Delta^i$ since it
appears in the phases of the Fourier transforms
of both $h^i_k(t-\Delta^i)$ and $h^i_j(t-\Delta^i)$ and cancels out when one is multiplied with
the complex conjugate of the other.
Since $h^i_k$ depends on 
just two orthogonal functions, $h_c$ and $h_s$, ($\langle h_c, h_s\rangle = 0$),  $\langle h_k^i,h_j^i\rangle$ depends
only on $\langle h_c, h_c\rangle$ and $\langle h_s,h_s\rangle$. These, in turn, do not depend on any of the remaining signal parameters since they all appear in the phase. Thus, $\langle h_c, h_c\rangle$ and $\langle h_s,h_s\rangle$ can be precomputed and stored. 
Evaluation of $\langle h^i_k(t-\Delta^i) | h^i_j(t-\Delta^i)\rangle$ for given $\alpha$ and 
$\delta$ then simply involves 
 taking algebraic combinations  of these stored scalars.

Next, 
transferring the detector dependent time shift $\Delta^i$ in $\langle x^i | h^i_k(t-\Delta^i)\rangle$  to the detector data,
\begin{align}
{\langle}x^i|h^i_k(t-\Delta^i){\rangle}^i &= {\langle}x^i(t+\Delta^i)|h^i_k(t){\rangle}^i. 
\label{eq:data_template_innprod}
\end{align}
replaces the cost of generating  $2D$  waveforms, namely, $h_c(t-\Delta^i)$ and $h_s(t-\Delta^i)$ for $i \leq D$, with that of doing $D$ time shifts. 

Finally,  we replace the data by over-whitened data, i.e., $\widetilde{x}^i(f) \rightarrow \widetilde{x}^i(f)/S_n^i(f)$, at the start of analysis. This saves the cost of division by $S_n^i(f)$ in constructing the integrand of the inner product in Eq.~(\ref{eq:innprod}) between data and templates.

\section{Particle swarm optimization}
\label{sec:pso}
The maximization over the intrinsic parameters in Eq.~(\ref{eq:FCASdetStat}),
\reviseMinorFive{namely $(\alpha, \delta, \tau_0,\tau_{1.5})$,} is carried 
out in \texttt{BINARIES} using PSO. By now the technical details of the PSO algorithm are well described in several papers (e.g., NMW), making the brief summary given below adequate for our purpose.

All PSO variants use the basic idea of evaluating the function to be maximized, called
the fitness function, at multiple locations, called particles, within a 
(bounded) search space. The particles move 
stochastically in the search space following rules, called dynamical equations, that 
implement a simple model of flocking behavior observed in bird swarms. In this model, the displacement (called velocity in PSO)
of each particle from one iteration to the next is affected by two 
forces, called social and cognitive, that attract the particle towards the best location  found by its neighbors and the best location found by the particle in its history. The iterations are initialized with random locations and 
velocities. While a variety of termination conditions  are available in the literature~\cite{engelbrecht2005fundamentals}, we use the simplest one where the 
algorithm is stopped after a specified number, $\niter$ of iterations. 

In this paper,  we use the same   PSO variant, called local best (lbest) PSO~\cite{bratton2007defining},
that was used for FCAS in NMW. In this variant, neighborhoods of particles are determined 
by the ring topology: particle
indices are arranged on a ring and a specified number of these
 on either side of a given index identify the neighbors of the corresponding particle. Specifically,   the total number of particles is set at $40$ with two neighbors for each particle. 
 In contrast, the variant used in WM was {\rm global best} (gbest) PSO where each particle has all other particles as its neighbors. 
For the same number of particles, the performance of lbest PSO as configured above has been demonstrated to be better for the FCAS search than gbest PSO.

In common with most practical stochastic optimization methods, PSO is not guaranteed,  even asymptotically,
to converge to the global maximum. As such, for a finite number of iterations,
one can only demand an acceptable 
probability of convergence to a specified region containing the global maximum.
One of the key elements behind the success of PSO in FCAS is the 
\reviseMinorTen{best-of-M-runs} strategy~\cite{birattari2007assess,mohanty2018swarm} for boosting the convergence
probability: multiple  runs of PSO,
utilizing independent random number streams, are performed on the same GW data and
the run that terminates with the best (maximum) value of $\Gamma^2(\Theta)$
provides both the coherent search statistic as well as the estimates of $\Theta$ and  $\Theta_{\rm ext}$. 

Along with $\niter$, the number of  runs, $\nruns$,
forms the only set of PSO parameters that are tuned in \texttt{BINARIES}. 
The metric used for tuning these parameters 
is based on the fact~\cite{Wang+Mohanty:2010}
that the global maximum of the coherent fitness
function should always be shifted away from the location,  $\Theta_0$,
of the true signal
parameters. This, after all,
is what leads to parameter estimation errors in the 
presence of noise. Consequently, the value of the coherent search statistic found 
by PSO, 
denoted as $\rho_{\rm coh}^\prime(\nruns,\niter)$, should at least be greater than 
the value, denoted as $\rho_{\rm coh}^{(0)}$, of the 
coherent fitness function at $\Theta_0$ if convergence to the global maximum is successful.
This leads to our definition of the tuning metric: 
\begin{equation}
   \mathcal{M}\left(\nruns,\niter\right)  =  {\rm Pr}\left(\rho_{\rm coh}^\prime(\nruns,\niter)< \rho_{\rm coh}^{(0)}\right)\!,
   \label{eq:tuning_metric}
\end{equation}
where ${\rm Pr}(A)$ is the probability of an event $A$. 
The goal of tuning is to bring $\mathcal{M}\left(\nruns,\niter\right)$ to an acceptably small value. Needless to say, this metric can only be estimated for simulated data where $\Theta_0$
is known. 

\section{runtime analysis}
\label{sec:comptime}

\texttt{BINARIES} is implemented in the C programming language and uses the \texttt{Intel MKL}
multi-threaded numerical algorithms library for computing FFTs.
Three nested parallelization layers are implemented. 
The outer 
layer uses \texttt{LAUNCHER}~\cite{Wilson:2014:LSF:2616498.2616534} to distribute 
independent runs of PSO across different nodes of a distributed computing cluster.
In the inner layer, a specified number of \texttt{OpenMP}~\cite{Dagum:1998:OIA:615255.615542} parallel processes (threads) evaluate 
PSO particle fitness values. Each process is further assigned 
a specified number of threads for use
by \texttt{MKL} functions. 

We have tested and compared the performance of \texttt{BINARIES}
on two different computing clusters, namely, Stampede 2 (S2) and  Lonestar 5 (LS5)
housed in the Texas Advanced Computing Center. The nodes used on S2 have one Intel Xeon Phi Knights Landing (KNL) processor per node with 68 cores supporting up to 4 threads per core.
The LS5 nodes used 
have two Intel Xeon {\sc E-5-2690} (Haswell) processors per node with 12 cores each supporting up to 2 threads per core.
While these
processors also differ in other details such as the clock rate and cache memory size that
are pertinent to computational speed, the main determinant for \texttt{BINARIES} is the number of concurrent \texttt{OpenMP} and \texttt{MKL} threads that can be supported.  
The threads related to fitness evaluations are distributed on all the processing cores of a node, with each node executing one PSO run.

The search space for PSO is typically taken to be a hypercube. Given that the search over the sky in \texttt{BINARIES} is not
partitioned, this translates into a rectangle in $\tau_0$,
$\tau_{1.5}$ space. To cover the entire range of $\tau_0$ and $\tau_{1.5}$ values
for CBC signals,  a set of
rectangles in this space 
along with their overlap fraction must be prescribed. Since the search in each rectangle 
can be conducted in parallel, and since the runtime -- defined as the wall-clock time taken to complete the analysis of a given segment of data -- does not depend on the size of the rectangle searched but on
 the number of detectors, data duration, number of PSO particles, and the number of PSO iterations, 
 having multiple rectangles does not increase the overall runtime of the PSO-based FCAS search. 
Similarly, the \reviseMinorTen{best-of-M-runs (c.f., Sec.~\ref{sec:pso})} strategy does not add to the runtime of \texttt{BINARIES} if all the runs are computed in parallel. 
Thus, for an analysis of runtime, it is sufficient to consider the search over only one rectangular region and one PSO run.

There are two factors that contribute to the computational speed of \texttt{BINARIES}: the improved
numerical implementation of the mathematical formalism described in Sec.~\ref{sec:efficientNumerics}
and, as described above, the number of parallel \texttt{OpenMP} and \texttt{MKL} threads used. 
We fix the number of \texttt{OpenMP} threads at $40$, the number of PSO particles used in \texttt{BINARIES}, while the number of 
\texttt{MKL} threads is varied to find the optimum operating point.

\begin{figure}
\centering
\includegraphics[scale=0.21]{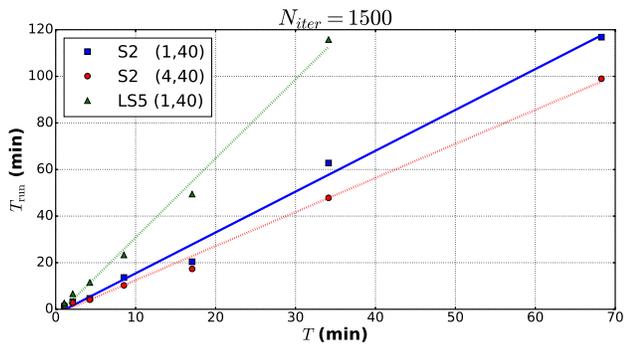}
\caption{\label{fig:comp_time_s2_vs_ls5}
Runtime, ($T_{\rm run}$),  as 
a function of the duration of data ($T$)  for \texttt{BINARIES} running on 
S2 and LS5 with different numbers of \texttt{MKL} threads. 
The number of \texttt{MKL} and \texttt{OpenMP} threads 
are shown as the first and second numbers, respectively, in the parentheses next to the name of the cluster (S2 or LS5). 
 The dependence of $T_{\rm run}$ on $T$ is approximately linear with slope  $1.75$, $1.46$, and $3.39$ for S2($1$,$40$), S2($4$,$40$), and LS5($1$,$40$), respectively. For large $T$, S2 is $\approx 2.3$ times faster than LS5. $T_{\rm run}$ in all the cases above is obtained for the number of PSO iterations set to $\niter=1500$.
}
\end{figure}

To quantify the effect of the improved numerical implementation 
and the use of \texttt{MKL} threads, we ran 
\texttt{BINARIES} on the same platform (LS5) as the codes used in NMW and found it to be $\approx 22$ times faster on average. 
The effect of different processors, and the different number of threads that they can support,
is shown in 
Fig.~\ref{fig:comp_time_s2_vs_ls5} where the runtime ($T_{\rm run}$)
of \texttt{BINARIES} 
 is analyzed
across S2 and LS5 as a function of the duration of data ($T$) and
different numbers of \texttt{MKL} threads. While, interestingly, 
the optimum number of \texttt{MKL} threads was found to depend on $T$ for short
data durations, its best value is
$4$ for the large $T$ of $60$~min considered in this paper. Overall, the 
change in processor from Haswell to KNL (with $4$ \texttt{MKL} threads)
provides a factor of $\approx 2.3$ speed up.  

\begin{figure}
\centering
\includegraphics[scale=0.22]{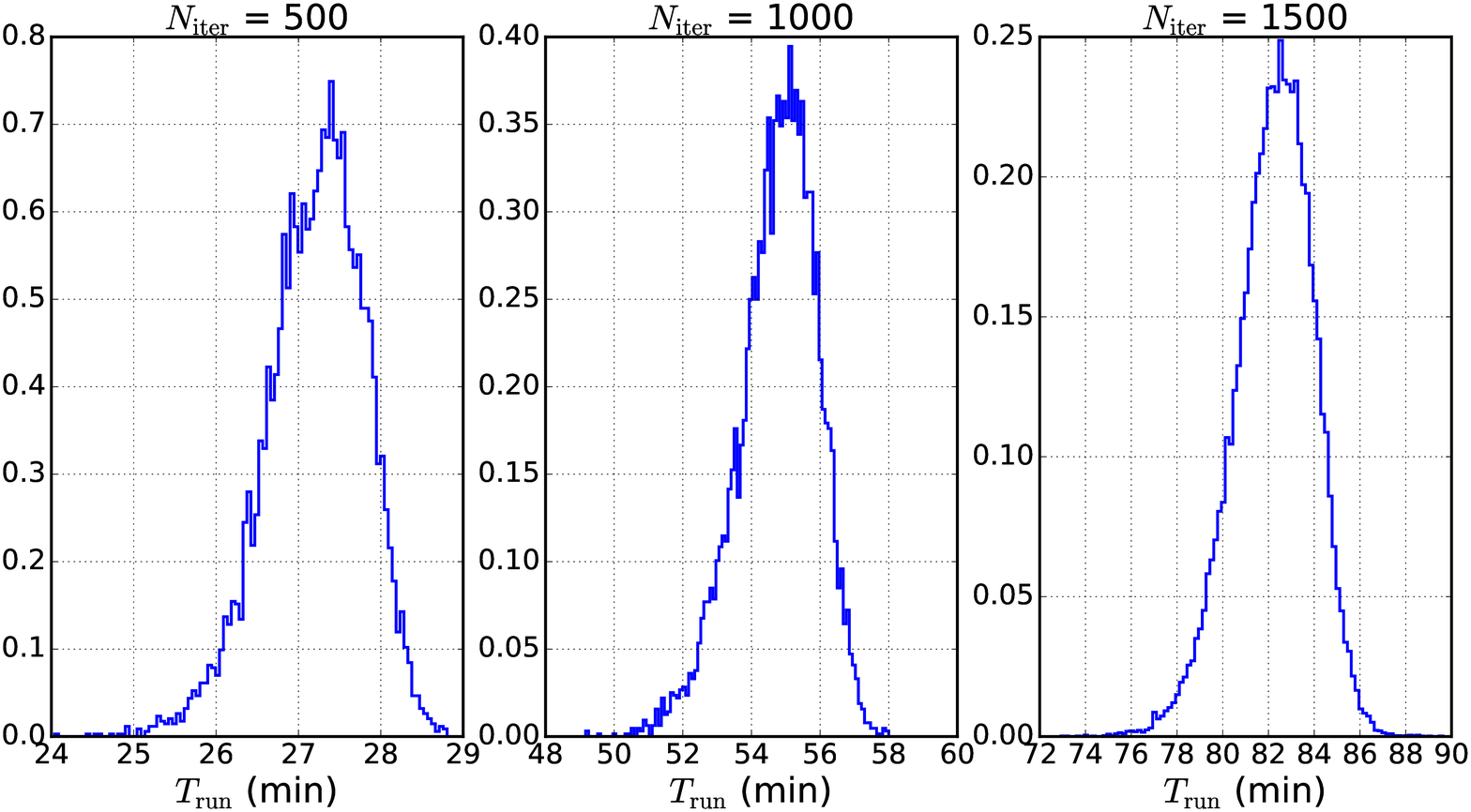}
\caption{\label{fig:binaries_tuning_computation_time_histograms} 
The distribution of runtime, $T_{\rm run}$, of \texttt{BINARIES} on the KNL processor  as a function of
the number, $\niter$, of  PSO iterations for duration of data $T=60$~min. The 1st, 50th, and 99th percentiles of the runtime (in minutes)  are $\{25.6,  27.3, 28.4\}$, $\{51.6, 54.9, 57.0\}$, and $\{77.7, 82.3, 85.7\}$ for
$\niter = 500$, $1000$, and $1500$, respectively. 
\texttt{BINARIES} was run with $40$ \texttt{OpenMP} and $4$ \texttt{MKL} threads in all cases.
Starting from the lowest $\niter$, the first two histograms were obtained from $7200$ trials while the third used $35200$ trials.
}
\end{figure}

Figure~\ref{fig:binaries_tuning_computation_time_histograms} shows the distribution of 
$T_{\rm run}$ on KNL and its dependence on the number, $\niter$,
of PSO iterations for $T=60$~min. We see that the runtime is quite stable, with a fairly narrow 
spread, and the average $T_{\rm run}$ has a linear dependence on $\niter$. 
\texttt{BINARIES} attains faster than real time processing speed, $T_{\rm run}< T$, for $\niter \leq 1000$. 

Besides the type of processor, the demands a code puts on other hardware parameters must
also be noted. Among these, the principal one is the available system random access 
memory (RAM). The RAM consumed
depends on  the sampling rate of the data, the use of single ($4$ bytes) or double 
($8$ bytes) precision data types, length of the data to be analyzed, number of detectors,  and most importantly, the number of concurrent PSO particle evaluations. For analysis in double precision of  $60$~min data sampled at $2048$~Hz, \texttt{BINARIES} requires a base amount of $576$~MB plus $192$~MB per detector for 
evaluating the fitness of a single PSO particle. 
For a total load of $40$ PSO particles, with concurrent evaluation of all fitness values,  the RAM required is $\approx 54$~GB for a four detector network.  For short data stretches of $60$~sec, keeping all other variables fixed, the RAM needed  is $\approx 900$~MB. 

\section{Results}
\label{sec:performance}
The performance of \texttt{BINARIES} is characterized using simulated data realizations.
Each realization consists of $60$~min long time series from the four-detector HLVK network at a sampling frequency of $2048$~Hz. 
The noise realizations in each time series are
generated following the model described in Sec.~\ref{sec:fcas}. 
To have a large number of noise realizations and yet remain within the cycle lengths 
of standard pesudo-random number generators, we adopt the standard approach used in GW searches 
for measuring background rates: unphysical and independent time shifts of $\geq 10$~sec
are applied to the noise 
time series from the different detectors to generate effectively 
new noise realizations. For our choice of 
the time shift and the number of time shifts, 1000 new realizations can be generated from a given data realization in this manner.

We use two different sky locations for generating strain responses: (L4) $\alpha = 32.09^\circ$,
$\delta = -53.86^\circ$, and (L5) $\alpha = 150.11^\circ$, $\delta = -60.16^\circ$. The 
polarization angle is set to be
$\psi = 30^\circ$ at both locations. L4 and L5 are two out of the $6$ locations used in WM and correspond to 
the 
best and worst condition numbers for the antenna pattern matrix 
${\bf F}(\alpha,\delta,\psi)$ of the HLVK network.

The binary component masses are kept equal and set to $1.5$~$M_\odot$. The signal time of 
arrival at the ECEF origin is set to be $20$~min after the start of the data and, for the 
chosen mass parameters and $f_\ast$, the strain response in each detector lasts $15.1$~min.

The signals are normalized to have a prescribed network signal to noise ratio (SNR) 
defined as
\begin{eqnarray}
    {\rm SNR} 
    & = & {\left[\sum_{i=1}^D \langle h^i | h^i \rangle^i \right]^{1/2}}\;,
    \label{optimalSNR}
\end{eqnarray}
For each location, we generate strain responses with ${\rm SNR}= 9$, $10$, $12$, and $15$. 

Results related to $H_0$ (noise-only) data are obtained from $1000$ realizations while $250$ $H_1$ (signal plus noise) 
data realizations  
are used for each combination of location and ${\rm SNR}$.  
For the search space of PSO,  $\alpha$ and $\delta$ cover the entire sky, while 
 $\tau_0 \in [500, 1500]$~sec and $\tau_{1.5}\in [5, 15]$~sec. 

In the remainder of this section, we first present the results for PSO tuning followed 
by the detection and estimation performance of \texttt{BINARIES}.
\subsection{PSO tuning}
The tuning procedure for PSO involves 
estimating the metric $\mathcal{M}\left(\nruns,\niter\right)$ defined in Eq.~(\ref{eq:tuning_metric}) from a set of 
simulated $H_1$ data realizations for a given combination of $\niter\in \{500, 1000, 1500\}$, $\nruns\in\{4, 8, 12\}$, and ${\rm SNR}$.  To reduce the 
computational burden involved in tuning, the number of $H_1$ data realizations 
is lowered to $120$ and only the L4 location is used. 

\begin{figure}
\centering
\includegraphics[width=0.45\textwidth,height=4in]{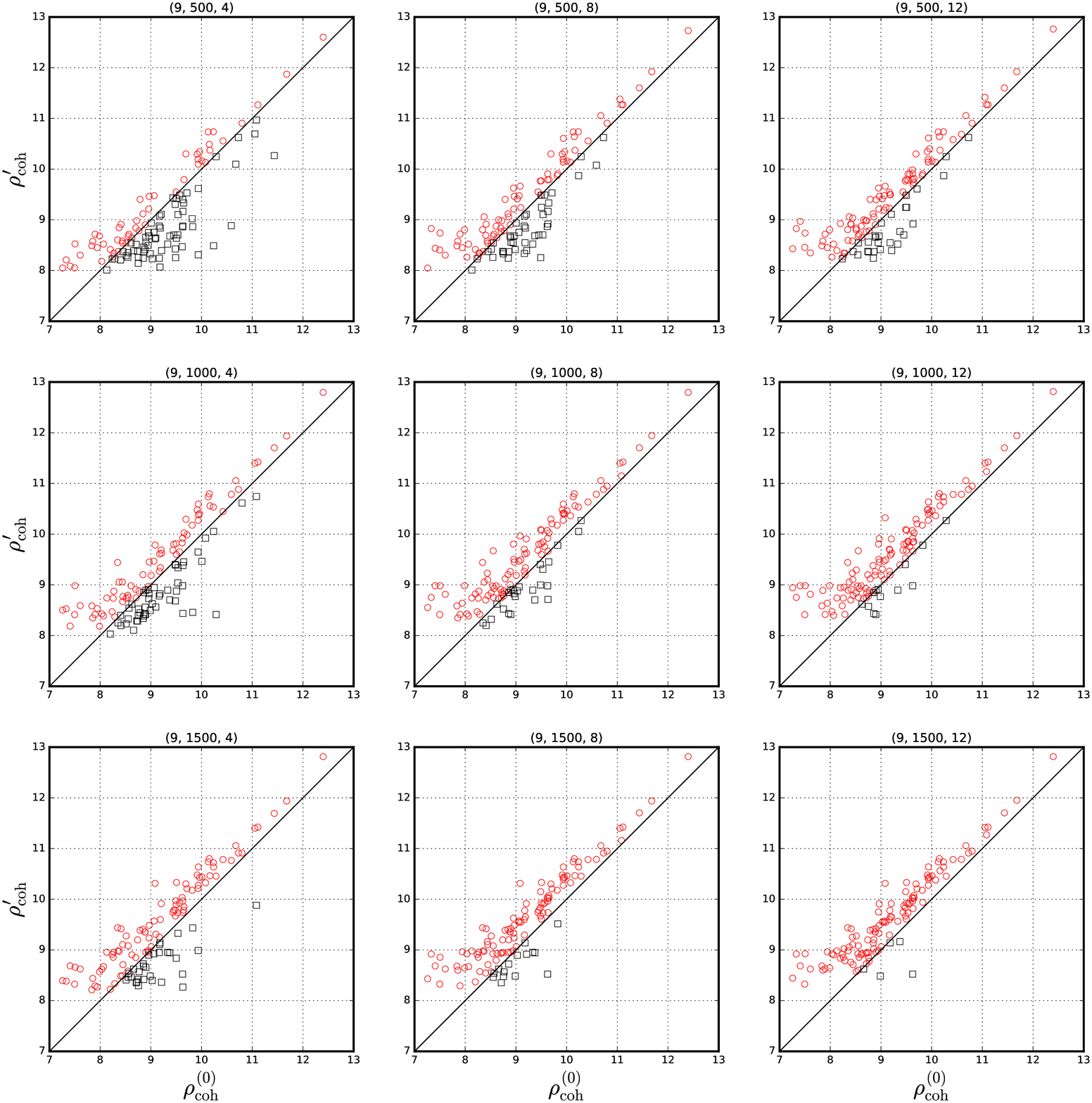}
\caption{\label{fig:binaries_tuning_css_s9}Scatterplots of the coherent search statistic found by PSO,
$\rho_{\rm coh}^\prime(\nruns,\niter)$, and the coherent fitness value at the true signal parameters, $\rho_{\rm coh}^{(0)}$, for an ${\rm SNR}=9$ source at the L4 location. Each subplot corresponds to the combination of $\niter$ and $\nruns$ stated in its title as (SNR,$\niter$,$\nruns$), and shows the values of $\rho_{\rm coh}^\prime(\nruns,\niter)$ and $\rho_{\rm coh}^{(0)}$ for $120$ data realizations. Points (black)
below the diagonal 
indicate instances in which PSO failed to converge to the global maximum of the coherent fitness function. 
}
\end{figure}

Figure~\ref{fig:binaries_tuning_css_s9}
illustrates how the performance of PSO evolves for a given SNR as $\niter$ and $\nruns$ are changed. 
Given a scatterplot from this figure corresponding to some $\niter$ and $\nruns$ combination, the simplest
estimate of $\mathcal{M}\left(\nruns,\niter\right)$ is just the fraction of points that fall below the diagonal.
This raw estimate can be improved upon using the bootstrap~\cite{efron1992bootstrap} based procedure 
introduced in NMW.

Table~\ref{Table:tuning} presents statistical summaries of the sample distribution of  $\mathcal{M}\left(\nruns,\niter\right)$ for different combinations of 
$\niter$, $\nruns$, and ${\rm SNR}$. 
We see that the tuning metric moves towards lower values, as desired, 
with an increase in $\niter$ or $\nruns$. Given the number of nodes and the number of threads per node in 
a distributed computing environment, the table entries can be used to find the appropriate values 
for $\niter$ and $\nruns$ to use. Similarly, given a desired value of the metric, the table allows us to
scope out the computing resources needed to achieve that value. 
\begin{table*}
\caption{ 
The PSO tuning metric $\mathcal{M}(\nruns,\niter)$ 
for a discrete set of 
 SNR values. For each $\nruns$ and $\niter$ combination, there are  
four rows corresponding (from top to bottom) to ${\rm SNR} = 9$, $10$, $12$ and $15$, respectively. In each row, the numbers from left to right are the $1$st and the $99$th percentiles of the sample distribution of 
$\mathcal{M}(\nruns,\niter)$, respectively.  
\label{Table:tuning}}
\centering
\begin{tabular}{| c | c | c | c | c | c | c | }
\hline
$\niter$ & $\nruns=2$ & $\nruns=4$ & $\nruns=6$ & $\nruns=8$ & $\nruns=10$ & $\nruns=12$\\
\hline
\hline
500 & $\begin{array}{lcr}0.633 & 0.783\\
\hline
0.633 & 0.792\\
\hline
0.325 & 0.500\\
\hline
0.042 & 0.150\\
\end{array}$
 & $\begin{array}{lcr}0.517 & 0.667\\
\hline
0.483 & 0.650\\
\hline
0.183 & 0.325\\
\hline
0 & 0.058\\
\end{array}$
 & $\begin{array}{lcr}0.433 & 0.583\\
\hline
0.408 & 0.558\\
\hline
0.125 & 0.242\\
\hline
0 & 0.033\\
\end{array}$
 & $\begin{array}{lcr}0.383 & 0.525\\
\hline
0.358 & 0.500\\
\hline
0.092 & 0.192\\
\hline
0 & 0.025\\
\end{array}$
 & $\begin{array}{lcr}0.342 & 0.483\\
\hline
0.325 & 0.450\\
\hline
0.075 & 0.167\\
\hline
0 & 0.017\\
\end{array}$
 & $\begin{array}{lcr}0.317 & 0.450\\
\hline
0.300 & 0.417\\
\hline
0.067 & 0.142\\
\hline
0 & 0.017\\
\end{array}$
\\
\hline
1000 & $\begin{array}{lcr}0.450 & 0.608\\
\hline
0.425 & 0.583\\
\hline
0.100 & 0.217\\
\hline
0 & 0.033\\
\end{array}$
 & $\begin{array}{lcr}0.308 & 0.467\\
\hline
0.292 & 0.433\\
\hline
0.033 & 0.117\\
\hline
0 & 0.008\\
\end{array}$
 & $\begin{array}{lcr}0.233 & 0.375\\
\hline
0.233 & 0.358\\
\hline
0.017 & 0.083\\
\hline
0 & 0.008\\
\end{array}$
 & $\begin{array}{lcr}0.192 & 0.317\\
\hline
0.200 & 0.317\\
\hline
0.008 & 0.067\\
\hline
0 & 0\\
\end{array}$
 & $\begin{array}{lcr}0.167 & 0.275\\
\hline
0.183 & 0.283\\
\hline
0.008 & 0.058\\
\hline
0 & 0\\
\end{array}$
 & $\begin{array}{lcr}0.142 & 0.250\\
\hline
0.167 & 0.258\\
\hline
0.008 & 0.050\\
\hline
0 & 0\\
\end{array}$
\\
\hline
1500 & $\begin{array}{lcr}0.333 & 0.492\\
\hline
0.308 & 0.467\\
\hline
0.042 & 0.133\\
\hline
0 & 0.017\\
\end{array}$
 & $\begin{array}{lcr}0.192 & 0.333\\
\hline
0.192 & 0.317\\
\hline
0.008 & 0.067\\
\hline
0 & 0\\
\end{array}$
 & $\begin{array}{lcr}0.133 & 0.250\\
\hline
0.133 & 0.250\\
\hline
0.008 & 0.042\\
\hline
0 & 0\\
\end{array}$
 & $\begin{array}{lcr}0.092 & 0.208\\
\hline
0.108 & 0.208\\
\hline
0.008 & 0.033\\
\hline
0 & 0\\
\end{array}$
 & $\begin{array}{lcr}0.075 & 0.175\\
\hline
0.092 & 0.183\\
\hline
0.008 & 0.025\\
\hline
0 & 0\\
\end{array}$
 & $\begin{array}{lcr}0.058 & 0.150\\
\hline
0.083 & 0.167\\
\hline
0.008 & 0.025\\
\hline
0 & 0\\
\end{array}$
\\
\hline
\end{tabular}
\end{table*}

Since the computing clusters available to us could easily accommodate 
the largest value of $\nruns$ in Table~\ref{Table:tuning}, we simply choose $\nruns = 12$ and 
look for the largest $\niter$ needed to achieve a sufficiently low value of $\mathcal{M}\left(\nruns,\niter\right)$.
Following this approach, we see that setting $\niter = 500$ for ${\rm SNR}\geq 12$ already 
gives a low value of $\leq 0.1$ for the first and $\approx 0.1$ for the $99$th percentile, respectively, of the tuning metric distribution. 
Thus, we do not need to move further down the table in this case. For ${\rm SNR}\leq 10$, on the 
other hand, one needs $\niter = 1500$ to meet similar  conditions. 
Thus, in summary, we choose $\nruns = 12$, $\niter = 500$, and $\nruns = 12$, $\niter = 1500$ 
for the analysis of data containing
${\rm SNR}\geq 12$ and ${\rm SNR} \leq 10$ signals, respectively. 

When analyzing real data, one would need to fix a target minimum ${\rm SNR}$ at which good performance of PSO 
is required and proceed in the same manner as above
to tune $\nruns$ and $\niter$. The training data for tuning in this
case could either be simulated or derived from a section of real data that is set aside for this purpose. \reviseMinorEight{There would be a single set of $\niter$ and $\nruns$ values when searching real data and a single detection threshold derived from the analysis of $H_0$ data with these parameters. However, for simulated data, where the signal SNR is known a priori, we can reduce computational costs by 
using $\niter$ and $\nruns$
tuned to different target minimum SNRs. For example, we can use the values tuned for a target minimum ${\rm SNR}=12$ instead of ${\rm SNR}=9$ to analyze data with ${\rm SNR}=15$ signals since this reduces the computational burden of the simulation without significantly affecting detection performance. }

Relating the tuned value of $\niter$ for ${\rm SNR}=9$  to the 
distribution of runtimes presented in
Fig.~\ref{fig:binaries_tuning_computation_time_histograms}, the runtime
of \texttt{BINARIES} for four-detector data on a KNL cluster 
can be expected to take $\approx 40\%$
longer than real-time. If the target minimum ${\rm SNR}$ is increased to $\gtrsim 12$, for which
$\niter = 500$,  \texttt{BINARIES} can analyze data at almost twice real-time speed with 
a $60$~min data segment processed in $\approx 27$~min on 
the average.

\subsection{Detection performance}
\label{sec:results_detection}

\begin{figure}
\centering
\includegraphics[width=\linewidth]{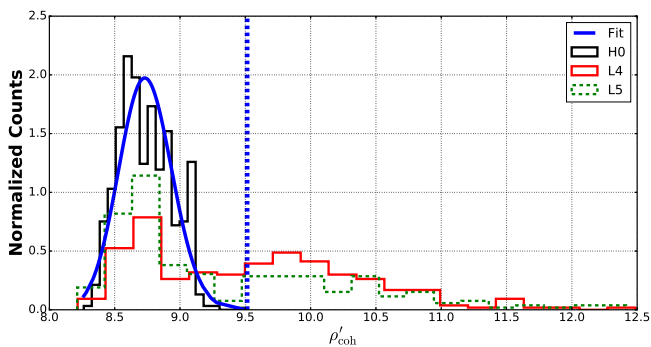}
\caption{\label{fig:css_hist_comparison_s9} Histograms of the coherent search statistic found by PSO, $\rho_{\rm coh}^\prime(12,1500)$, under the  (black curve) $H_0$  and (red and green curves) $H_1$  hypotheses for ${\rm SNR} = 9$.
Under $H_1$, the histograms corresponding to
the source
locations (red) L4 and (green) L5 are shown separately.  Also shown (solid blue curve) is the best-fit lognormal probability density function for the $H_0$ distribution. The dashed line at $\rho_{\rm coh}^\prime=9.5$ marks the detection threshold  obtained from the best-fit for a false alarm rate of $1$~event per year. 
The \pvalue of the two-sample KS test between the $H_1$ samples is $7.2\times 10^{-5}$.  
Retaining only $\rho^\prime_{\rm coh} > 9.0$ results in a \pvalue of $0.33$.
}
\end{figure}

\begin{figure}
\centering
\includegraphics[width=\linewidth]{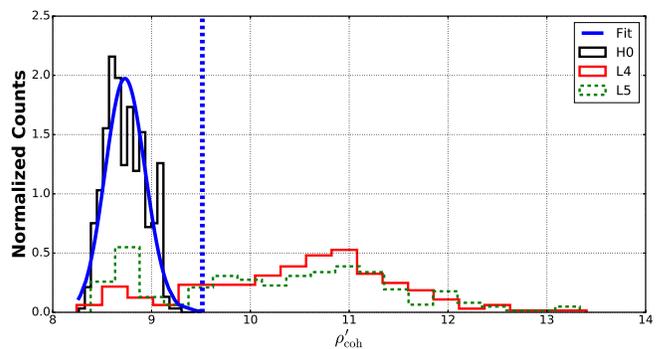}
\caption{\label{fig:css_hist_comparison_s10}Same as Fig.~\ref{fig:css_hist_comparison_s9} except that
${\rm SNR} = 10$ and
the \pvalue of the two-sample KS test is $0.02$. Retaining only $\rho^\prime_{\rm coh} > 9.0$ results in a \pvalue of $0.83$. 
}
\end{figure}

\begin{figure}
\centering
\includegraphics[width=\linewidth]{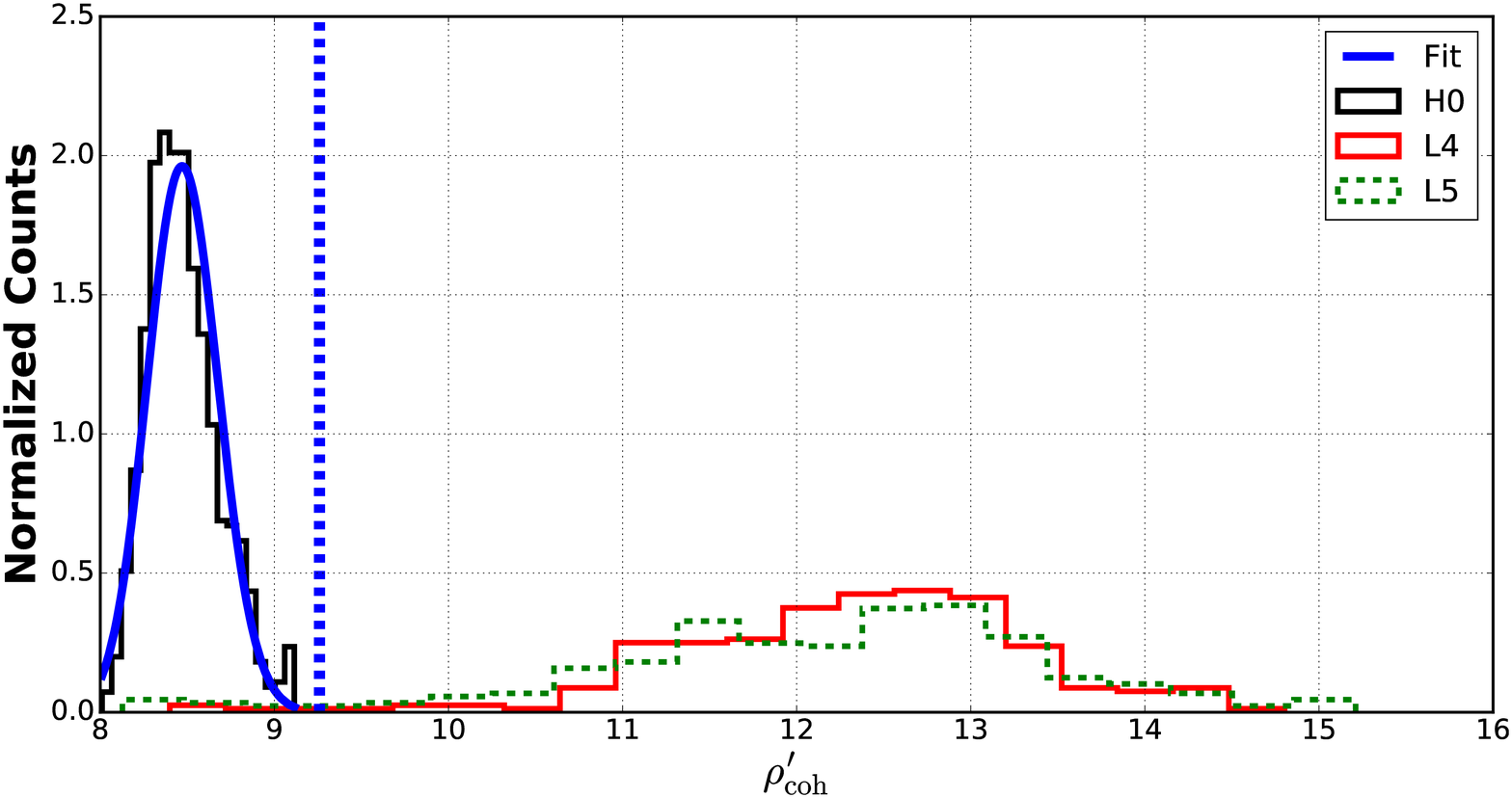}
\caption{\label{fig:css_hist_comparison_s12_n500_m12}
Histograms of the  coherent search statistic found by PSO, $\rho_{\rm coh}^\prime(12,500)$, under the  (black curve) $H_0$  and (red and green curves) $H_1$  hypotheses for ${\rm SNR} = 12$.
Under $H_1$, the histograms corresponding to
the source
locations (red) L4 and (green) L5 are shown separately.  Also shown (solid blue curve) is the best-fit lognormal probability density function for the $H_0$ distribution. The dashed line at $\rho_{\rm coh}^\prime=9.3$ 
marks the detection threshold  obtained from the best-fit for a false alarm rate of $1$~event per year.
The \pvalue of the two-sample KS test between the $H_1$ samples is $0.23$.  
Retaining only $\rho^\prime_{\rm coh} > 9.0$ results in a \pvalue of $0.48$.
}
\end{figure}

\begin{figure}
\centering
\includegraphics[width=\linewidth]{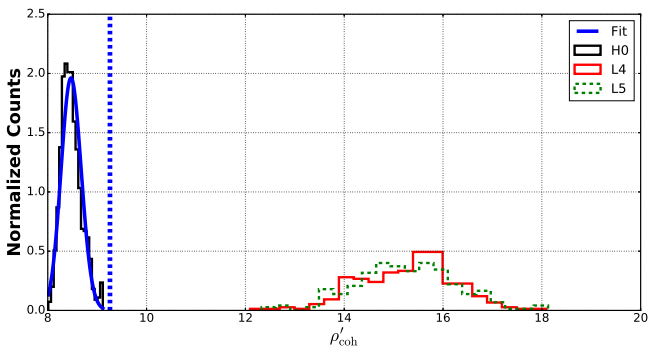}
\caption{\label{fig:css_hist_comparison_s15_n500_m12}
Same as Fig.~\ref{fig:css_hist_comparison_s12_n500_m12} except that 
${\rm SNR} = 15$ and the \pvalue of the two-sample KS test is $0.52$.
There is no change in the \pvalue when retaining only $\rho^\prime_{\rm coh} > 9.0$ because all the
values satisfy this cutoff.
}
\end{figure}

Figures~\ref{fig:css_hist_comparison_s9} to~\ref{fig:css_hist_comparison_s15_n500_m12} show
the estimated distributions of the coherent search statistic found by PSO, $\rho^\prime_{\rm coh}$,
under the $H_0$ and $H_1$ hypotheses for 
different ${\rm SNR}$ values. The $H_0$ distributions in each figure are obtained using the values of $\niter$ and $\nruns$ tuned, as described above,
for the respective SNR. The distribution under $H_1$ is shown separately for the two 
source locations, L4 and L5.  

A comparison of the $H_1$ histograms at a given ${\rm SNR}$ shows that the 
distribution of the coherent search statistic is principally governed by ${\rm SNR}$.
However, the condition number of the antenna pattern matrix, ${\bf F}(\alpha,\delta,\psi)$, at the source location also has an effect. This is indicated by 
performing a two-sample 
Kolmogorov-Smirnov (KS) test of the null hypotheses that the two $H_1$ samples are drawn from the 
same probability distribution. The \pvalue of the 
test -- the probability of obtaining the observed KS statistic value under the null hypothesis -- is listed in the captions of the figures. It suggests a
clear difference between the probability distributions at
${\rm SNR}\leq 10$ but a statistically insignificant
difference for ${\rm SNR}\gtrsim 12$. It is interesting to note that 
for the shorter data length ($64$~sec) used in NMW, the two distributions did not differ 
significantly even for ${\rm SNR}=9$. 

The two-sample KS test was also carried out between the $H_1$ samples after excluding values
of $\rho^\prime_{\rm coh}< 9.0$. Under this restriction, the \pvalue in all cases indicates that there is no statistically significant difference between the two samples. All of the difference clearly arises from the distribution of  $\rho^\prime_{\rm coh}$ at low values. Since the cut-off value of $9.0$ is below any 
reasonable detection threshold (see below), this implies that the distribution of $\rho^\prime_{\rm coh}$ values for detected signals will always be observed to be independent of GW source location (for at least ${\rm SNR}\geq 9.0$ tested here). 

The distribution of $\rho^\prime_{\rm coh}$ under $H_0$ is observed to change for 
different values of $\niter$, with a shift towards lower values for smaller $\niter$.
This can be understood qualitatively from the fact that, unlike the case of 
$H_1$, the fitness function under $H_0$ has many local maxima dispersed throughout the search space with comparable heights. The likelihood of PSO missing the global maximum is, therefore, higher 
in this case as its exploration ability, which is controlled by $\niter$, is reduced. Consequently,
lower $\rho^\prime_{\rm coh}$ values are found by PSO under $H_0$ for lower $\niter$.
This effect is quite mild, however, leading to only a marginal change in the detection threshold.

For a fiducial false alarm rate (FAR) of 1~false event per year, the false alarm probability
per $60$~min data segment is $1.14\times 10^{-4}$ if there is
no overlap between consecutive segments.
The corresponding detection threshold, obtained by fitting a log-normal distribution to 
the $H_0$ histograms, for $\niter = 1500$ and
$\nruns =12$, is $\eta = 9.5$. The same procedure for the $H_0$ histogram under 
$\niter = 500$ and $\nruns =12$ yields a slightly lower threshold of $9.3$.

Table~\ref{Table:detection} shows detection probabilities 
for different combinations of ${\rm SNR}$ and sky locations under the conservative choice
of using the higher detection threshold.
The effect of antenna pattern condition number is clear: for low ${\rm SNR}$ values, 
the detection probability at L5 is markedly worse than at L4. The effect of the condition number  and,
consequently, the discrepancy in detection probability dissipates for ${\rm SNR}\gtrsim 12$.

The necessarily finite and typically small sample size used for tuning PSO means that 
the true value of the metric $\mathcal{M}\left(\nruns,\niter\right)$ need not be zero even when
its estimated value from the sample is.
For a 
sufficiently large number of trials, therefore, there will always be a finite fraction
in which the coherent search statistic value
found by PSO, $\rho_{\rm coh}^\prime$,
drops below the coherent fitness at the true location, $\rho_{\rm coh}^{(0)}$. However, 
while each such dropout is an instance 
of failure to converge to the global maximum of the coherent fitness function, what matters is the loss in the detection probability that this entails. Detection probability is reduced
by a dropout event
only if  $\rho_{\rm coh}^{(0)}$
exceeds the detection threshold but $\rho_{\rm coh}^\prime$ does not. To measure this effect, we define the loss in detection 
probability,
\begin{eqnarray}
L_{\rm DP} & = & P(\rho^\prime_{\rm coh}\leq \eta | \rho^{(0)}_{\rm coh}\geq \eta )\;,
\label{eq:loss_detprob}
\end{eqnarray}
where $P(A|B)$ denotes the conditional probability of event $A$
given event $B$, and $\eta$ is the detection threshold.
An examination of the estimated $L_{\rm DP}$ values shown in Table~\ref{Table:detection} 
shows that it decreases quite rapidly as ${\rm SNR}$ increases, becoming too small to 
measure with our simulations for ${\rm SNR}=15$. As with the detection probability, we have also
shown $L_{\rm DP}$ for each location separately, and it is evident that again the condition 
number has a major effect with L5 showing a significantly higher loss in detection probability. In fact, the loss in detection probability at ${\rm SNR}=9$ arises entirely from the L5 data realizations.
\begin{table}
\caption{
Detection probabilities for all combinations of ${\rm SNR}$ and sky locations at a detection threshold of $\eta = 9.5$
corresponding to a FAR of $\approx 1$
false event per year. 
Also listed (third column) is the loss in detection probability, $L_{\rm DP}$,
defined in Eq.~(\ref{eq:loss_detprob}). The last two columns separate
the contributions to 
$L_{\rm DP}$ by the two sky locations used for generating data realizations.
\label{Table:detection}}
\centering
\begin{tabular}{| c | c | c | c | c | c |}
\hline
 ${\rm SNR}$ & L4 & L5 & $L_{\rm DP}$ & $L_{\rm DP}$/L4 & $L_{\rm DP}$/L5 \\
\hline
\hline
$9$ & 0.504 & 0.372 & $11.798\%$ & $0.000\%$ & $21.429\%$\\
\hline
$10$ & 0.820 & 0.720 & $4.324\%$ & $0.538\%$ & $8.152\%$\\
\hline
$12$ & 0.992 & 0.972 & $1.613\%$ & $0.806\%$ & $2.419\%$ \\
\hline
$15$ & 1.0 & 1.0 & $0.000\%$ & $0.000\%$ & $0.000\%$\\
\hline
\end{tabular}
\end{table}

A feature of the distribution of the coherent search statistic
under $H_1$ that may appear surprising at first 
is the appearance of a bump, seen as an excess in 
histogram counts, at low values. This excess is reduced (see Fig.~\ref{fig:css_hist_comparison_s10}) and 
ultimately disappears (cf. Figures ~\ref{fig:css_hist_comparison_s12_n500_m12} and~\ref{fig:css_hist_comparison_s15_n500_m12}) as the signal
${\rm SNR}$ goes up. The appearance of this bump is simply 
due to the coherent search statistic being the global maximum of
the coherent fitness function, not its value at a fixed location. 
The presence of a signal only affects the distribution of 
coherent fitness function values in a small region of the full
parameter space, the distribution elsewhere being close to that for $H_0$ data. For a sufficiently weak signal, the probability that 
the global maximum escapes from the small region affected by the signal is higher. When this happens, the value of the coherent search statistic is drawn from its distribution under $H_0$. For a strong signal, on the other hand, the global maximum stays confined to the region close to the signal and its governing distribution is that 
under $H_1$. In other words, the distribution of the coherent search statistic in the presence of a signal is actually a mixture of two distributions with the probability of sampling from either depending on the strength of the signal. A counterpart of this effect is also
seen in Fig.~\ref{fig:binaries_tuning_css_s9}, where a leveling off
is observed in the values of $\rho_{\rm coh}^\prime$ at low 
values of $\rho_{\rm coh}^{(0)}$.

\subsection{Estimation performance}
\label{sec:estimation}
\begin{figure}
\centering
\includegraphics[scale=0.45]{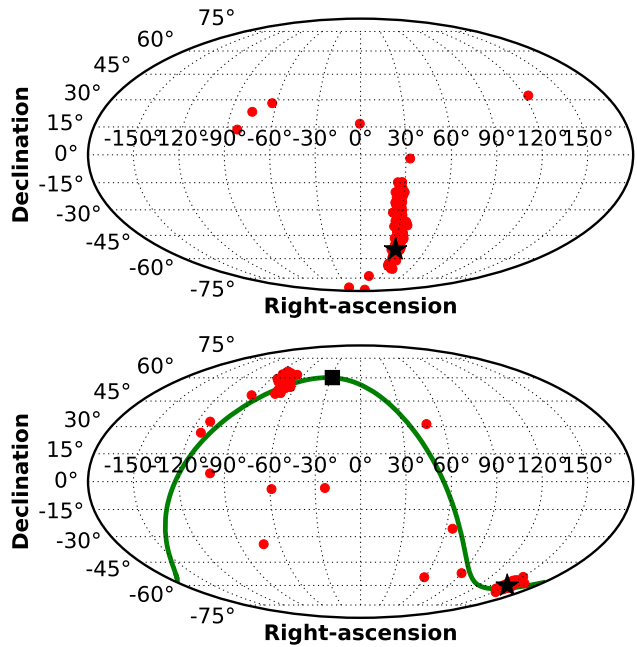}
\caption{\label{fig:binaries_skymap_mollweide_M3L4S12_M3L5S12}
Estimated sky location of a ${\rm SNR}=12$ source at (top) L4 and (bottom) L5 locations. In both plots the true location of the source is shown by a star, and estimated locations from 
$H_1$ data realizations are shown by red circles. In the bottom plot, the square shows the antipode of the true location.
}
\end{figure}

\begin{figure}
\centering
\includegraphics[scale=0.45]{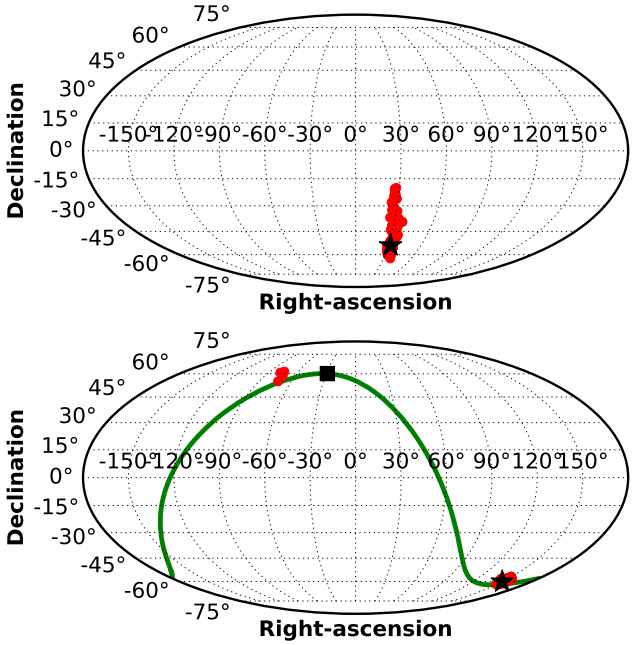}
\caption{\label{fig:binaries_skymap_mollweide_M3L4S15_M3L5S15}
Same as Fig.~\ref{fig:binaries_skymap_mollweide_M3L4S12_M3L5S12} but for SNR=15.
}
\end{figure}

The parameter estimation performance of \texttt{BINARIES} is characterized here for 
${\rm SNR} \in \{12, 15\}$ for which the corresponding detection probabilities (cf. 
Table~\ref{Table:detection}) are near unity. Figures~\ref{fig:binaries_skymap_mollweide_M3L4S12_M3L5S12} and~\ref{fig:binaries_skymap_mollweide_M3L4S15_M3L5S15} show the distribution of the 
estimated sky locations at these ${\rm SNR}$ values. They follow the same pattern as seen in 
NMW for shorter signals: while the estimated locations of a source at L4 are fairly 
well-clustered around the true value, two distinct clusters appear for a source at L5. The 
secondary cluster of estimated locations in the latter case appears close to the point 
that is antipodal to the true location. This is a clear manifestation of the 
condition 
number of the antenna pattern matrix ${\bf F}(\alpha,\delta,\psi)$, which is worse at L5 than L4,
and consistent with its effect on detection probabilities at lower ${\rm SNR}$ values. While
its deleterious effect on detection probability disappears for ${\rm SNR} \in \{12, 15\}$, 
it remains in force for sky localization error. 

It is important to emphasize here that, as
was determined during the tuning process,  PSO almost always converges to the global maximum
for ${\rm SNR}\in \{12, 15\}$. This indicates strongly that
the appearance of the secondary cluster of locations 
is not due to 
a failure in convergence to the global maximum but its actual
jump to that location due to the effect of noise.  The exact mechanism by which the 
condition number creates a secondary location is yet to be elucidated and is left to future work.

\begin{figure}
\centering
\includegraphics[scale=0.42]{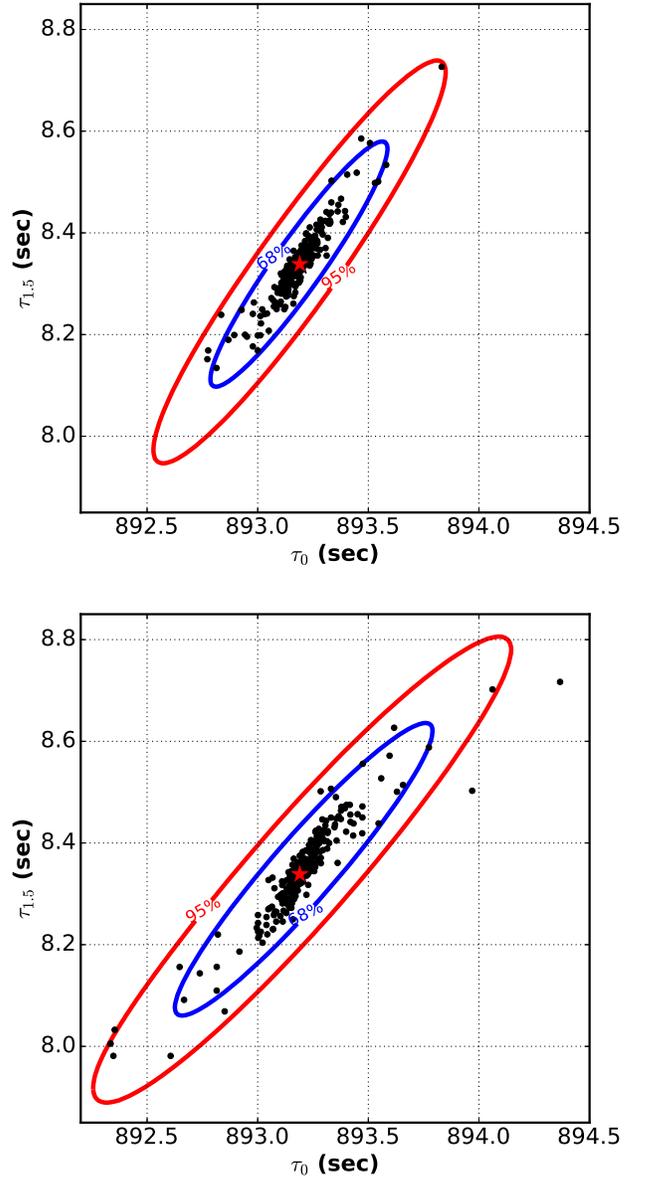}
\caption{\label{fig:binaries_chirp_map_s12}
Estimated chirp times for a ${\rm SNR}=12$ source at (top) L4 and (bottom) L5 locations. In both plots the true values of the chirp times are marked by a star, and estimated values from 
$H_1$ data realizations are shown by black dots. The contours shown include (blue) $68\%$ and (red) $95\%$ of the total probability of a kernel density estimate, obtained using 
a Gaussian kernel with a bandwidth of 2,  of the 2-dimensional 
probability density function. 
}
\end{figure}

\begin{figure}
\centering
\includegraphics[scale=0.42]{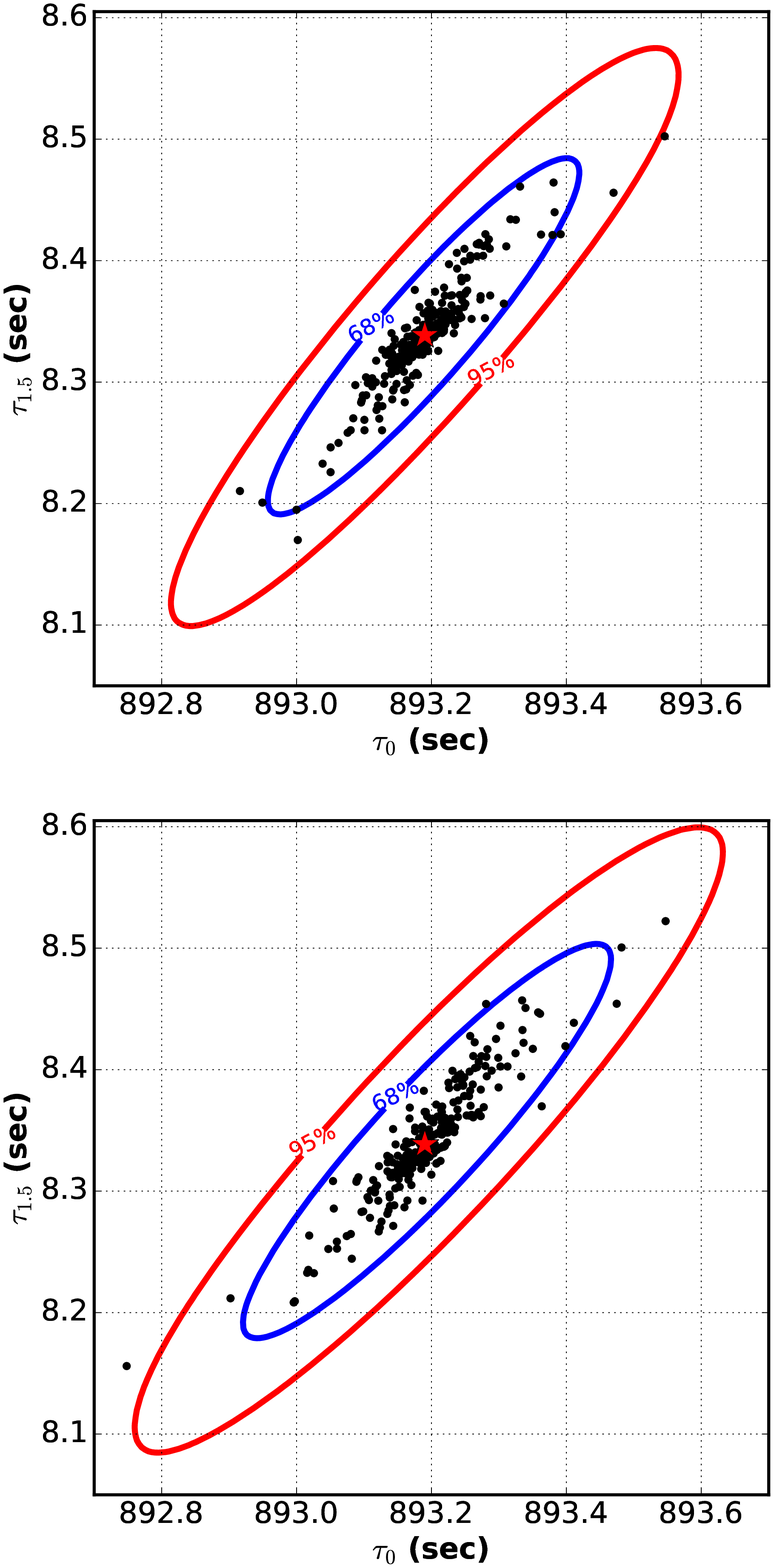}
\caption{\label{fig:binaries_chirp_map_s15}
Same as Fig.~\ref{fig:binaries_chirp_map_s12} except
 ${\rm SNR}=15$.
}
\end{figure}

 Figures \ref{fig:binaries_chirp_map_s12} and~\ref{fig:binaries_chirp_map_s15} show the 
 distribution of estimated chirp time parameters corresponding to sources at L4 and L5. 
 The condition number of the antenna pattern matrix does not have any noticeable effect on
 these distributions. This is consistent with the known result, from CRLB analysis, that 
 errors in sky location have low correlation with errors in the chirp times~\cite{singer2016rapid}.

\section{Discussion}
\label{sec:discussion}

We have presented performance results for a new code, called \texttt{BINARIES}, that
implements a PSO-based FCAS search. Using this code we were able to verify
that \reviseMinorNine{the PSO-based search, which had previously been demonstrated for few minute long
data segments in WM~\cite{weerathunga2017performance} and NMW~\cite{normandin2018particle}, can be extended to the analysis of the longest data length needed for CBC searches}. 

\reviseOne{The \texttt{BINARIES} code differs significantly from the code used previously in NMW: (i) A new and more efficient numerical implementation of the mathematical formalism underlying FCAS search is used, and (ii) a third parallelization layer is introduced using a multi-threaded numerical algorithms library.  Together, these
changes result in a factor of $\approx 22$ enhancement in the speed of the code.}

Simulated $60$~min long data realizations from a network of four 
second-generation detectors (HLVK) at their respective design sensitivities 
were generated for different source locations and 
${\rm SNR}$ values. The mass parameters of the source were chosen to be $(1.5, 1.5)$~$M_\odot$, which is representative of long duration CBC signals. 
Depending on the target minimum ${\rm SNR}$, the runtime of \texttt{BINARIES} 
ranges from $40\%$ slower 
than real time to about twice as fast. For the latter, it is possible to
detect ${\rm SNR}\gtrsim 12$ signals with a detection probability of nearly unity
at a FAR of $\approx 1$~false event per year. This is the first demonstration that the computational 
barrier confronting an always-on FCAS search,
which promises substantially better
sensitivity than semi-coherent searches, is not insurmountable at astrophysically 
relevant signal strengths.

Analysis of $H_0$ (noise-only) data shows that  ${\rm SNR}=9$ falls below
the detection threshold required by the FAR chosen in this paper. This could be an intrinsic 
feature of a four detector network or the result of PSO failing to converge often enough 
to the global maximum of the coherent fitness function. For the current choice of 
PSO variant, a realistic signal strength that allows confident detections
is ${\rm SNR}\approx 10$ for which the 
detection probability is $\gtrsim 70\%$. 

\reviseThree{It should be emphasized here that the network SNR values above correspond to a four-detector network. In the simple case where a signal has equal single-detector SNR across a network, an ${\rm SNR}=10$ corresponds to a two-detector ${\rm SNR}=(10/\sqrt{4})\times \sqrt{2}=7.07$, which is below the network SNR of events reported in  current LIGO-only semi-coherent searches.  Moreover, the detection probability for a given SNR and number of detectors also depends on the mass range of the search. In NMW, where a high mass binary with a total mass of $29$~$M_\odot$ was used as an example, a detection probability of $\approx 70\%$, at a false alarm rate of $1$~false event per year, was achieved for a lower four-detector ${\rm SNR}=9$.}

Our study of sky localization error 
highlights the important effect of the condition number of the antenna
pattern matrix, which measures the ill-posedness of the GW network analysis problem arising from the 
relative orientations of detectors.
It was found that the error region on the sky can split into two widely 
separated areas for a source 
location having a high condition number. The detection probability for such a source is also
reduced although this is 
a significant effect only for ${\rm SNR}\lesssim 10$. 
The loss in detection probability, caused by the failure of PSO to converge to the global maximum, is consistently higher for
the source location having a higher condition number. 
The above results 
suggest
that incorporating some form of regularization~\cite{klimenko:2005,2006CQGra..23.4799M,2006CQGra..23S.673R} 
in the derivation of the coherent search statistic for CBC signals is important for improving 
sky localization and detection sensitivity. 

The number of parallelization layers used in \texttt{BINARIES} can be increased by offloading the bulk of the computations involved in fitness evaluation, such as the inner product of arrays in Eq.~(\ref{eq:data_template_innprod}), to Graphics Processing Units (GPUs). Given that a 
larger number of MKL threads leads to a significantly faster processing speed, the $O(10^3)$
threads available on GPUs promise to provide an even greater improvement. This investigation is currently in progress.

\reviseTwo{While \texttt{BINARIES} has been applied to simulated data in this paper, our eventual goal is to apply it to O1, O2, and future open data. This requires embedding \texttt{BINARIES} in an end-to-end search pipeline that includes data conditioning, glitch vetoes, post-processing, and background rate analysis. Results 
from the analysis of real data following the completion of the full pipeline will be reported in future papers.}

\section*{Acknowledgements}
The contribution of S.D.M. to this paper was supported by National Science Foundation (NSF)
 grant PHY-1505861. 
We acknowledge the Texas Advanced Computing Center (TACC) at The University of Texas at Austin for providing HPC resources that have contributed to the research results reported within this paper.
URL: www.tacc.utexas.edu. 

\bibliography{ms.bbl}

\end{document}